\documentclass[12pt,a4papper]{article}
\usepackage[spanish]{babel}
\usepackage[latin1]{inputenc}

\usepackage{graphics}

\usepackage{amssymb}
\usepackage{color}
\usepackage{amsmath}
\usepackage{amsfonts}

\newcommand{\bs}{\bigskip}

\newcommand{\beq}{\begin{equation}}
\newcommand{\eeq}{\end{equation}}
\newcommand{\beqa}{\begin{eqnarray}}
\newcommand{\eeqa}{\end{eqnarray}}
\newcommand{\beqn}{\begin{eqnarray}}
\newcommand{\eeqn}{\end{eqnarray}}
\renewcommand{\r}{{\bf r}}

\newcommand{\B}{{\bf B}}

\newcommand{\A}{{\bf A}}
\newcommand{\C}{{\bf C}}
\newcommand{\E}{{\bf E}}

\newcommand{\dif}{{\rm d}}
\newcommand{\R}{{\bf R}}

\renewcommand{\v}{{\bf v}}

\newcommand{\medio}{\frac{1}{2}}

\begin{document}

\begin{center}

{\Large \bf A Topological Structure in the Set of Classical Free}

\medskip

{\Large \bf  Radiation Electromagnetic Fields}

\vspace{3mm}

{\large A. F. Ra\~nada$^1$ and A. Tiemblo$^2$}

\medskip

{10 July 2014}

\end{center}

\vspace{2cm}

{\LARGE \bf Contents}

\vspace{5mm}

{\bf 1. Introduction}

\vspace{3mm}

{\bf 2. The Hopf index and the Faraday 2-form}

\vspace{3mm}

{\bf 3. Whitehead's Theorem}

\vspace{3mm}

{\bf 4. Field equations of the topological electromagnetic fields}

\vspace{3mm}

{\bf 5. Electromagnetic helicity and the discretization of the energy of the fields}

\vspace{6cm}

$^1$ Dept. of Applied Physics III, University Complutense, Plaza de las Ciencias, 28040 Madrid, Spain. Tel: 34 629475028.

$^2$ Institute of Fundamental Physics, Consejo Superior de Investigaciones Cient\'ificas, Serrano 113b, 28006 Madrid, Spain, Tel 34 600764231. 

\newpage


\pagestyle{myheadings} \markright{\footnotesize \bf
Ra\~nada-Tiemblo, 10 July 2014}

\centerline{\bf Abstract}

\bs

The aim of this work is to proceed with the development of a model
of topological electromagnetism in empty space, proposed
by one of us some time ago and based on the existence of a
topological structure associated with the radiation fields in
standard Maxwell's
 theory. This structure consists in pairs of complex scalar fields, say  $\phi$ and $\theta$, that
can be interpreted as maps $\phi,\theta: S^3\mapsto S^2$, the level lines of which are
 orthogonal to one another, where $S^3$ is the compactified physical
3-space $R^3$, with only one point at infinity, and $S^2$ is the
2-sphere identified with the complete complex plane. These maps
were discovered and studied in 1931 by the German mathematician H.
Hopf, who showed that the set of all of them can be ordered in
homotopy classes, labeled by the so called Hopf index, equal to
$\gamma=\pm 1,\,\pm 2,\,\cdots ,\, \pm k,...$ but without $\gamma=0$.
In the model presented here and at the level of the scalars $\phi$
and $\theta$, the equations of motion are highly nonlinear;
however there is a transformation of variables that converts
exactly these equations (not by truncation!) into the linear
Maxwell's ones for the magnetic and electric fields $\B$ and $\E$.

\section{Introduction}

 The purpose of this paper is to continue the development of a model
of topological electromagnetism, proposed  by one of us in three
papers in the years 1989-1992  \cite{Ran89}-\cite{Ran92} and based on several curious and intriguing similarities between
some topological ideas of the German mathematician H. Hopf  \cite{Hop31} and of ordinary
Maxwell's electromagnetism.
In addition to these previous works, a paper published in Nature Physics in 2008 by W. T.
Irvine and D. Bouwmeester attracted the attention of some
mathematical physicists towards the topological
 aspects of classical electromagnetism \cite{Bou08}. They stated in the first lines of their paper: ``Maxwell's equations allow for curious solutions characterized by
 the property that {\it all} electric and magnetic  field lines are closed loops with {\it any} two electric or
 (magnetic) lines are linked. These little known solutions, constructed by Ra\~nada,
are based on the Hopf fibration" (see also
\cite{Ran95b}-\cite{Ran12}). Since 2008, an increasing number of
papers are dedicated to this subject. It does happen that the
classical free radiation electromagnetic fields ({\it i. e.} the subset of the electromagnetic fields
verifying $\E\cdot \B=0$ and $|\E|=|\B|$)\  \cite{Enk13} are
associated with time dependent pairs of maps $\phi,\theta:
S^3\mapsto S^2$, with orthogonal level lines,
where $S^3$ is the compactified physical 3-space with only one
point at infinity, that is $R^3\cup \infty$, $S^2$ is the 2-sphere identified with the
complete complex plane and $\phi$ and $\theta$
 are two  scalar complex fields. This results in several thought-provoking connections
  between electromagnetism and topology. In order to understand better these ideas,
  we will make use of an interesting theorem discovered in 1947 by the British mathematician Whitehead \cite{Whi47}.

At first sight, however, the idea of similarities between the works of Hopf and Maxwell might seem strange. Indeed, there is an infinity of time dependent free electromagnetic fields, while the time does not intervene in the expression of Hopf's invariants. Nevertheless, we show in this work that it is possible to interpret  some of the quantities that appear in
Hopf's theories as Cauchy initial time values, particularly his work on what is now called  ``the Hopf invariant"\  or ``the Hopf index". Moreover we take here, in addition to a differential one, an algebraic point of view which simplifies the arguments.

This association of the works of Maxwell and Hopf has surprising
consequences. Indeed, it turns out that, in the model proposed here: \\
(i) The free radiation
 electromagnetic fields can be classified in
homotopy classes labeled by the values of the corresponding Hopf
indexes $\gamma =\pm 1,\pm 2,\ldots \pm n\ldots$, which
are topological constants of the motion. Note that $\gamma =0$ would imply zero energy.\\
(ii) The
magnetic and electric lines are the level curves of the the maps
$\phi$ and $\theta$, respectively; consequently each line  can be
labeled by a complex number \cite{Ran89}-\cite{Ran92}, \cite{Enk13}. \\
 (iii) At the level of the
maps $\phi$ and $\theta$, the equations of motion are highly
nonlinear; however they can be transformed  exactly into the
linear Maxwell's equations for the fields $\E$ and $\B$. \\
(iv) Both the magnetic and electric  level lines of any pair of fields
 are linked with linking
number $\ell=\gamma$.

Moreover an interesting property of our model is that the electromagnetic energy of the fields is discretized.
The words ``quantum"\   and ``discrete" (or
``quantization"\ and ``discretization") will refer here
to quantum and topological effects, respectively.\\

To ensure that all the dynamical quantities have the
correct dimensions, the model makes use of the basic constant $a=\hbar c\mu_0$, the product of  the Planck constant, the speed of light and the permeability of empty space. We will use space and time  dimensionless coordinates $(X,Y,Z,T)$, related to the physical ones $(x,y,z,t)$
of the SI as $(X,Y,Z,T)=\lambda (x,y,z,ct),\quad \lambda^2 r^2=\lambda^2(x^2+y^2+z^2)=X^2+Y^2+Z^2=R^2,$  where $c$ is the speed of light  and $\lambda$ is a  constant inverse length that characterizes the size of the knot (its inverse gives a measure of the mean quadratic radius of the energy distribution of an electromagnetic field). In this work and for simplicity, we will take $c=1$ and $\lambda =1$.

Note that these ideas refer to a branch of mathematics, the topology, that most probably will play  a major role  in the future of fundamental physics.
As the renowned mathematician M. Atiyah stated ``Both quantum physics and topology lead from the continuous to the discrete" \cite{Ati88,Ati90}.
 Strictly speaking this is nothing new, think if not of Dirac's monopole, the Aharonov-Bohm effect, superconductivity,  superfluidity, etc, but there is a novelty now in the growing  number of  physicists  that work on the role of topology in classical electromagnetism \cite{Bou08,Enk13},\cite{Bes09}, \cite{Tho98}-\cite{Arr10}, \cite{Bes09}, \cite{Wol58},
 often in a way related to fluid mechanics \cite{Ric96}-\cite{Bar01} and with support from mathematicians  \cite{Ati88,Ati90,Fad97}, \cite{Bot82,Kuz80,Kho14}.

 \section{The Hopf index and the Faraday 2-form}

 In this work, we will consider pairs of smooth complex scalars fields, say $\phi (\r )$ and $\theta(\r)$, one valued
at infinity.  In this case, these  two fields
 can be interpreted as maps $\phi, \theta: S^3\mapsto S^2$,
after identifying, via stereographic projection, $R^3\cup
{\infty}$ with the sphere $S^3$ and the complete complex plane
$C\cup {\infty}$ with the sphere $S^2$. The word ``map" without
further qualification must be understood in this work as ``map  $S^3\mapsto S^2$".

Let us consider a brief summary of Hopf's ideas.  Given a smooth  map $\psi: S^3\mapsto S^2$, the inverse image  of any point $c'$
in $S^2$ is generically a closed curve, that can be labeled by the complex number $c'$.
In the case of a pair of points or, equivalently,  a pair of complex numbers $a$ and $b$ in the sphere $S^2$, $\psi^{-1}(a)$ and $\psi^{-1}(b)$, labeled also as $a$ and $b$,
are in general two closed lines that can be linked or not. Let us assume that they are linked and let $\Sigma_a$
be an oriented surface bordered by $\psi^{-1}(a)$, noted as $\partial \Sigma_a=\psi^{-1}(a)$. This closed curve $\psi^{-1}(a)$  is the set of preimages of the point $a$ in $S^2$.
The line $\psi^{-1}(b)$ intersects this oriented surface  $\ell$ times, taking into account in the sign of the intersection.
This number $\ell$ is equal to the degree of the restriction of the map $\psi$ to $\Sigma_a$. The degree of such a map is equal to the integral over $\Sigma_a$ of the pull-back of $\sigma$, the normalized area 2-form of $S^2$. If expressed in stereographic coordinates, this 2- form  is
\begin{equation} \Psi=\frac{1}{2\pi i}\frac{\dif \bar{\psi}\wedge \dif \psi}{(1+\bar{\psi}\psi)^2},\label{1.10}\end{equation}
where the bar over a complex number means complex conjugation.

 According to the De Rham theorem \cite{Nas82} and because $\Psi$ is closed in $S^3$, the second group of cohomology of which is trivial, there must exist a 1-form $\cal A$ such that $\Psi=\dif {\cal A}$. Hence, the Hopf index can be expressed as
\begin{equation} \gamma=\int_{S^3}{\cal A}\wedge \Psi. \label{1.40}\end{equation}

This number $\gamma$ is equal to the linking number of any pair of lines and
does not depend on the particular pair, since by moving
continuously from a position into a different one the inverse images can
neither untie nor tie any further to one another, otherwise there would be a point in $S^3$ which at a time would be the inverse image of two separated points in $S^2$ or, in other words, this point would have two images in $S^2$.

Let $f: S^3\mapsto S^2$. The linking number $\ell$ of two lines $f^{-1}(a)$ and $f^{-1}(b)$ is generically equal to the area of the surface $\Sigma _a$  if measured with the pull-back to $S^3$ of $\sigma$ and both are equal to the Hopf index $\gamma$, so that $\gamma= \ell= \pm 1,\,\pm 2,\,\pm 3... ,\,\pm k$..., so that both can be equal to any integer number, excluding 0.  This is why, when Hopf published his theory, some people said that he had shown that the space can be completely filled with circumferences (in the topological sense). It must be added that any pair of such closed curves is a link. The Hopf index is then $\gamma=\ell$.

The map
\begin{equation} \phi (X,Y,Z)=\frac{(X+iY)}{(Z+i(R^2-1)/2)}, \label{1.50}\end{equation}
was proposed by Hopf himself,  as an example in which his index is non-zero, in fact it is $\gamma=1$. The system of its level curves is the Hopf fibration.
 The same can be said of its complex conjugate
$\bar{\phi}=(X-iY)/(Z-i(R^2-1)/2)$), its index being $\gamma=-1$. Note that all the level lines
of $\phi({\bf R})$  thread once the circumference $Z=0$, $R=1$,
turn once around the $Z$ axis and any pair of them is linked once.
On the other hand, the pair of lines $\phi=0$ and
$\phi=\infty$ that are the $Z$-axis and the circumference $Z=0$,
$R=1$ are a couple of linked limes.

 We will see now why Faraday's conception of the force lines suggests a topological structure for electromagnetism.
In his view, forged along many hours of laboratory work, the force lines were  real and tangible, a revealing indication
 that something very especial occurred along them, some kind of perturbation of the space of a nature still not well  understood at his time.
 Let us follow his original view, representing the dynamics of the electromagnetic fields in empty space by the evolution of their magnetic
and electric lines; in other words, let us attempt to develop what could be called a line-dynamics. In order to do that, we need
to deduce the electromagnetic tensor from the expressions of these lines. Starting with a static situation and as a simple tentative idea, we could represent
the magnetic lines by the equation $\phi(X,Y,Z)=\phi_0$, where $\phi$ is a complex function and $\phi_0$  a complex constant labeling each line. However, we are searching a more realistic model, so that we will include the time and take the magnetic lines as  the level curves of $\phi(X,Y,Z,T)$, which define a line at any point  for each value of time . As the magnetic field is tangent to them, it can be written as $\B=g\nabla \bar{\phi}\times \nabla \phi$ (bars over complex numbers indicate complex conjugation), where $g$ is a well behaved function which, because $\nabla \cdot \B=0$, depends on ($\r, t$) only
through $\phi$ and $\bar{\phi}$.
\begin{equation} \B=g(\phi, \bar{\phi})\nabla \phi \times \nabla \bar{\phi}. \label{Tem.60}\end{equation}
We can then do the same for the electric field, introducing a second scalar $\theta$, so that
\begin{equation} \E= -g(\theta,\bar{\theta})\nabla \theta\times \nabla\bar{\theta},\label{Tem.70}\end{equation}
with the condition $\E\cdot \B=0$. Note that we accept here that the same function $g$
appears in the two equations (\ref{Tem.60}) and (\ref{Tem.70}).
This explains the statement made in the first paragraph
of the introduction that the electromagnetic fields of the model
are built on two scalars. However, $\theta$ can be deduced from
$\phi$ through a duality relation so that these
fields have just two degrees of freedom. This point
 will be clarified in the following.

As explained before, the scalars $\phi$ and $\theta$
can be interpreted as  maps $S^3\mapsto S^2$. Maps
of this kind have nontrivial topological properties so that {\em
the attempt to describe electromagnetism by the evolution of the
force lines, represented as the level curves of two complex
functions, suggests in a convincing  way the
appearance of a topological structure}. And an interesting one as will be
seen.

The previous considerations suggest that there might be a formal relation between the sets of pairs of maps
$\phi, \theta: S^3 \mapsto S^2$ and the set of radiation electromagnetic fields  in empty space \cite{Ran92,Ran89,Ran90,Ran92b,Bou08}. Or between Hopf's and Maxwell's theories.
However, this possibility seems surprising at first sight, because of the great difference between these two grand creations of Hopf and Maxwell:
 the time plays no role in the Hopf theory, while it is very important in electromagnetism. Nevertheless, we will be able to solve this apparent
 contradiction assuming that the solutions obtained in Hopf theory are the initial value at time $t_0$ of some time dependent functions. Moreover and because of evident covariance
reasons, they can be extended to four dimensional space-time, as in
eq. (\ref{1.10}), in such a way that we have now
the area form $\Phi$ as well as the tensor $\Phi_{\mu\nu}$
\begin{equation}   {\Phi}=\frac{1}{2\pi \mbox{i}}   \frac{\dif \bar{\phi}\wedge \dif \phi} {(1+\bar{\phi}\phi)^2}, \qquad
 \Phi_{\mu\nu}=\frac{1}{2\pi {\rm i}}
\,\frac{\partial_\mu\bar{\phi}\partial_\nu\phi-\partial_\nu\bar{\phi}
\partial _\mu\phi}{(1+\bar{\phi}\phi)^2}.   \label{Tem.90}\eeq

All these argument show that it could be a thought provoking idea to attempt to build a topological model of electromagnetism (TME) in empty space.  If the coordinates are rotated cyclically ($X,Y,Z)\rightarrow  (Y,Z,X) \rightarrow (Z,X,Y)$, the following three maps are obtained
\begin{equation} \phi=\frac{2(X+iY)}{2Z+i(R^2-1)},\quad \theta=\frac{2(Y+iZ)}{2X+i(R^2-1)},\quad \varphi=\frac{2(Z+iX)}{2Y+i(R^2-1)}.\label{M.10}\end{equation}
It is easy to see that the these three maps have the same Hopf index, equal to 1. The three fibrations are mutually orthogonal at each point and any two of the three maps form a pair of dual maps from which  an electromagnetic knot can be built. A we will see, the lines of the fibrations are the {\it magnetic lines}, {\it the electric lines} and the {\it lines of the energy flux}, tangent this one to the Poynting vector. This is a nice mathematical structure.

\section{Whitehead's theorem}

The British mathematician J. H. C. Whitehead  discovered in 1947 an integral expression for the Hopf invariant, or index,  which is important for our purposes \cite{Whi47}.
Let us consider a twice differentiable complex map $f:S^3\mapsto S^2$, where $S^2\subset R^3$ and $S^3\subset R^4$ are spheres, $S^2$ having unit radius. Let $x^1,x^2,x^3$ be local coordinates
 for $S^3$ and $\lambda,\mu$ be local coordinates for $S^2$ and let $\sigma(\lambda,\mu)$ be the area density on $S^2$, {\it i. e.} what multiplies $\dif \lambda \wedge \dif \mu$ in the expression
 of the area 2-form on $S^2$.
The functions $\lambda,\,\mu$ depend on $(x^1,x^2,x^3)$ which allows to express locally the map $f$.

Whitehead started from the
 expression
\begin{equation} u_{ij}= \frac{1}{4\pi}\sigma(\lambda,\mu)\,\frac{\partial(\lambda,\mu)}{\partial(x^i,x^j)}, (i,j=1,2,3)\label{2.10},\end{equation}
where  $\lambda,\,\mu$ stand for the functions
$\lambda(x^1,x^2,x^3),\,\mu(x^1,x^2,x^3)$. The only condition on $\lambda$ and $\mu$ is that $\eta =\sigma(\lambda,\mu)\dif \lambda \wedge \dif \mu$ be the area form on $S^2$.

The $u_{ij}$ are the components of an alternating tensor in $S^3$,
which happens to have some common properties with the electromagnetic tensor
 in three dimensions $F_{ij}$. As shown by Whitehead the divergence
of this tensor vanishes, what is to say that
\begin{equation} \partial u_{23}/\partial x^1+\partial u_{31}/\partial x^2+\partial
u_{12}/\partial x^3=0.\label{2.20}\end{equation}
It turns out that the vector ${\bf b}= (u_{23},\,u_{31},\,u_{12})$ has much in common with a magnetic field. In fact, according to the De Rham's theorem  \cite{Nas82} a covariant vector field
$(v_1,v_2,v_3)$  exists such that
\begin{equation}\partial v_i/\partial x^j-\partial v_j/\partial x^i=u_{ij},\label{2.30}\end{equation}
where $u_{ij}$ is similar to a magnetic field and $v_j$ plays, in this theory, a similar role as the vector potential. As this  paper shows, this relation between electromagnetism and topology has deep consequences.
Whitehead proved then, after a long calculation,  that
\begin{equation}
\frac{1}{4\pi}\,\int_{S^3}\sigma\left|\begin{array}{ccc}v_1&v_2&v_3\\
\partial_1\lambda &\partial_2\lambda &\partial_3\lambda\\
\partial_1\mu &\partial_2\mu
&\partial_3\mu
\end{array}\right|\,\dif x^1\dif x^2 \dif x^3 = \gamma,\label{2.40}\end{equation}
where $\gamma$ is the Hopf invariant of the map $f$. This is Whitehead's formula. It gives the Hopf index as the integral over $S^3$ of a determinant times the area density $\sigma(\lambda,\mu)$ on $S^2$. Note that the functional dependence on $\lambda$ and $\mu$ on $x^1,x^2,x^3$ determines the map $f$ and the value of $\gamma$.

\section{Field equations of the  topological electromagnetic fields}

 The purpose of this work is to find fields with two characteristics: (a) to satisfy Maxwell's equations in empty space and
(b) to have topological properties. It happens that the following
expressions  comply with these two conditions, if  $(\phi,\theta)$
are two maps $S^3\mapsto S^2$  and the level lines of the two maps
are orthogonal to one another \beq \Phi _{\mu\nu}  =\frac{1}{2\pi
i}\,\frac{\partial _\mu\bar{\phi} \partial _\nu \phi - \partial
_\nu\bar{\phi} \partial _\mu \phi}{(1+\bar{\phi}\phi)^2},\quad
\Theta _{\mu\nu}  =\frac{1}{2\pi i}\,\frac{\partial
_\mu\bar{\theta} \partial _\nu \theta - \partial _\nu\bar{\theta}
\partial _\mu
\theta}{(1+\bar{\theta}\theta)^2}.\label{Tem.110}\eeq
 Note that $(\Phi_{\mu\nu}, \Theta_{\mu\nu})$ are formally similar to the electromagnetic tensor $F_{\mu\nu}$.
 Making now the following change of variables
$\phi = \rho e^{iq}$ we obtain
\begin{equation}\Phi _{\mu\nu}={(\partial_\mu f\partial _\nu q -
\partial_\nu f\partial _\mu q),\quad \mbox{ with }\quad f= -1/(2\pi(1+\rho^2)), \label{Tem.120}}\end{equation}
 where $f$ and $q$ are
known as the Clebsch variables. It must be stressed that

 (\ref{Tem.120}) can be written as \beq \Phi_{\mu\nu}= \partial
_\mu A_\nu - \partial _\nu A_\mu,\quad \mbox{ with }\quad  A_\mu =
f\partial _\mu q \label{TEM.140}\eeq
 Behold that $f\partial_\mu q$ is not a total derivative, otherwise it would be a pure gauge.

The simplest  Lagrangian analogous to the electromagnetic one, is
\beq {\cal
L}=-\frac{1}{4}\Phi_{\mu\nu}\Phi^{\mu\nu},\label{Tem.150}\eeq the
Lagrange equations of which are $\partial_\mu
\Phi^{\mu\nu}\partial _\nu f=0$ and $\partial_\mu
\Phi^{\mu\nu}\partial _\nu q=0.$ In order to simplify the algebra,
we introduce the variable $z_\nu =\partial ^\mu \Phi_{\mu\nu}$, so
that we can write the three equations \beq z_\nu\partial ^\nu f
=0,\quad z_\nu\partial ^\nu q=0, \quad \partial ^\nu z_\nu
=0.\label{Tem.160}\eeq The two first equations are the Lagrange
ones and the third is trivial. This is important because it will
allow us to prove that the tensor $\Phi_{\mu\nu}$  satisfies
Maxwell's equations. Note that $z_\nu=0$ are in fact the second
pair of Maxwell equations.

A comment could be in order. The same motion equations are obtained if we add to the ordinary electromagnetic Lagrangian a constraint fixing $A_\mu= f\partial_\mu q$. The value of the Lagrange multipliers is fixed by the motion equations, so that both theories are, in fact, the same. Furthermore, in spite of the very complex non linear appearance  of the field equations, we will see that, if written in terms of  $f$ and $q$, using only structural properties, their solutions can be classified in terms of three simple classes, as will be shown immediately.

Let us take now the equation
 \beq z^\beta\, ^*\Phi_{\beta \alpha}= \partial^\rho\, \Phi_\rho\,^\beta\,^*\Phi_{\beta\alpha}=
  \partial^\rho(\Phi_\rho\,^\beta\, ^*\Phi_{\beta\alpha})- \Phi^{\rho\beta}\partial _\rho\, ^*\Phi_{\beta\alpha}.\label{Tem.170}\eeq
It turns out that the term between parentheses vanishes, so that
 \begin{eqnarray}\Phi_\rho^\beta\,^*\Phi_{\beta\alpha} &=& \frac{1}{2}\,
(\partial_\rho f\partial _\beta q-\partial _\beta f\partial _\rho
q)\epsilon_{\beta\alpha\gamma\delta}\partial _\gamma
f\partial_\delta q \nonumber
=\epsilon_{\beta\alpha\gamma\delta}\partial_\rho f\partial _\beta
q\partial _\gamma f\partial _\delta q
 -\epsilon_{\beta\alpha\gamma\delta}\partial _\beta f\partial _\rho q\partial_\gamma f\partial _\delta q = 0.\\ \label{Tem.180}\end{eqnarray}
  We have thus
\begin{equation} z_\beta\,^*\Phi^\beta\, _\alpha
=-\Phi^{\rho\beta}\partial _\rho\, ^*\Phi_{\beta \alpha}
=-\medio\, \Phi^{\rho\beta}(\partial _\rho\,^*\Phi_{\beta
\alpha}-\partial_\beta\,^*\Phi_{\rho \alpha}) =- \medio
\Phi^{\rho\beta}(\partial _\rho\, ^* \Phi_{\beta
\alpha}+\partial _\beta\,^*\Phi_{\alpha\rho})=0 \label{Tem.190}
\end{equation}
Using in (\ref{Tem.190}) the cyclic identity, we obtain
   \beq z^\beta\,^*\Phi_{\beta \alpha}=\medio
\Phi^{\rho\beta}\partial
_\alpha\,^*\Phi_{\rho\beta}=\medio\epsilon^{\rho\beta\gamma\delta}(\partial_\rho
f\partial_\beta q-\partial _\rho q\partial _\beta
f)\Phi^{\rho\beta}\partial _\alpha(\partial _\gamma f\partial
_\delta q-\partial_\gamma q\partial _\delta f)=0,\label{Tem.200} \eeq

This equation can be written also as
$\epsilon_{\alpha\beta\gamma\delta} z^\beta\partial ^\gamma
f\partial^\delta q=0$. Multiplying by
${\epsilon^\alpha_{\beta'\gamma'\delta'}}$
it is found, that
\beq \quad (z_\beta\Phi_{\delta\gamma}+z_\delta\Phi_{\gamma\beta}+z_\gamma\Phi_{\beta
\delta}) \partial _\gamma f\partial _\delta q=0.\label{Tem.210}\eeq

Using the motion equations (\ref{Tem.160}) we get finally
$$z_\beta[(\partial f \partial q)^2-(\partial f)^2(\partial q)^2] =0,$$
where
$$[(\partial f \partial q)^2-(\partial f)^2(\partial q)^2]=-\medio \Phi_{\mu\nu} \Phi^{\mu\nu}=|\E|^2-|\B|^2,$$
where, as will be seen soon in the following, $\B$ and $\E$ are the magnetic and the electric fields, respectively.
There are three possible cases \\
\begin{equation} a)\; E^2-B^2=0,\; z_\beta =0,\qquad
 b) \;E^2-B^2\neq 0,\; z_\beta = 0,\qquad
 c)\; E^2-B^2=0,\;z_\beta \neq 0.\label{Tem.230}\end{equation}
We are interested in this paper in radiation solutions of Maxwell equations, so that we will consider only the case a).

 Let us define
\beq \Phi_{\mu\nu}=\partial_{[\mu}f\partial_{\nu]}q\quad \mbox{
and }\quad
\Theta_{\mu\nu}=\partial_{[\mu}g\partial_{\nu]}p\label{Tem.240}\eeq
and impose their relation as \beq \Phi_{\mu\nu}= \medio
\epsilon_{\mu\nu\rho\sigma}\Theta^{\rho\sigma},\quad \mbox{ and
}\quad \Theta_{\mu\nu}= \medio
\epsilon_{\mu\nu\rho\sigma}\Phi^{\rho\sigma}\label{Tem.250},\eeq
that expresses  the duality property in the sense of the
interchange between $\B$ and $\E$.
It follows that \\
 $$z_\nu =0, \mbox{ because }  \partial ^\mu\Phi_{\mu\nu}=\medio
\epsilon_{\mu\nu\rho\sigma}\partial ^\mu\Theta ^{\rho\sigma}=0.$$
From the first (\ref{Tem.240}) and the first (\ref{Tem.250}) we get
\begin{eqnarray} &&\Phi_{12}=\partial_3g\partial _0p-\partial _0g\partial _3 p, \nonumber \\
&&\!\!\!-\Phi_{13}=\partial _2g\partial _0p-\partial_0g\partial _2p\label{Tem.280}\\
&&\Phi_{23}=\partial _1g\partial
_0p-\partial_0g\partial_1p,\nonumber \end{eqnarray}

$$\partial_0g=-\frac{\Phi_{12}\partial_2g+\Phi_{13}\partial _3g}{\Theta _{23}}, \qquad
\partial_0p=-\frac{\Phi_{12}\partial_2p+\Phi_{13}\partial _3p}{\Theta _{23}}$$

Substitution in the third equation (\ref{Tem.280}) eliminates the
time derivatives, so that the following relation is valid for all
time \beq   \Phi_{12}\Theta ^{12}+ \Phi_{13}\Theta
^{13}+\Phi_{23}\Theta ^{23}=\B\cdot \E=0,\label{Tem.320}\eeq
where $(\Phi_{12},\,\Phi_{23},\,\Phi_{31}) =\B$ and  $(\Theta_{12},\,\Theta_{23},\,\Theta_{31})=\E.$
Obviously the same is obtained using  the second equations in (\ref{Tem.240}) and (\ref{Tem.250}).

Taking now the square of the equation (\ref{Tem.250}) one gets finally
\beq\E^2-\B^2=0.\label{Tem.340}\eeq

 As we said before the time is absent from the Hopf theory. Nevertheless, because the equations
 (\ref{Tem.320}) and (\ref{Tem.340}) are algebraic equations, they are valid for arbitrary values of the time. In fact they can be adopted as the Cauchy conditions for $t=0$, see the details in \cite{Ran01}.

\section{The electromagnetic helicity and the discretization of the energy of the fields}

Whitehead's formula (see section 3) can be written in terms of the two coordinates $\lambda$ and $\mu$ plus the vector $\v$, as follows
\begin{equation}\frac{1}{4\pi}\int_{S^3}\sigma \, {\bf v}\cdot (\nabla \lambda \times \nabla \mu)\,\dif x^1\dif x^2 \dif x^3 =\gamma. \label{2.140}\end{equation}
It is clear that $\nabla \cdot (\nabla \lambda \times \nabla \mu)=0$ and $\nabla \times \v= (\nabla \lambda\times \nabla \mu)$, so that $(\nabla \lambda \times \nabla \mu)$ and $\v$  are  similar to the magnetic field $\B$ and its vector potential $\A$, respectively.
Let us take the coordinates $\lambda
=\cos \vartheta$ (polar) and $\mu =2\varphi$ (azimuth) which are adequate for our purpose.  The
area 2-form in $S^2$ is then $\eta=\sigma(\lambda, \mu) \dif
\lambda\wedge \dif \mu$, with $\sigma (\lambda,\mu)=1$ so that $\int _{S^2}\eta =\int_{S^2}\dif \lambda\wedge
\dif \mu = 4\pi$. It turns out that $\sigma =1$ in this example and will be so in the applications of the rest of this work.

Up to now, we have used dimensionless quantities but, in order to apply these ideas to electromagnetism, it is convenient to introduce the dimensions.
Changing the variables in equation (\ref{2.140}) by introducing the multiplicative constant $\sqrt{4\pi/ a}$ so that $\v= \sqrt{4\pi/a}\,\A$,  $\nabla \lambda \times \nabla \mu=\sqrt{4\pi/a}\,{\bf B}$ (and $\v'= \sqrt{4\pi/a}\,\C$,  $\nabla \lambda' \times \nabla \mu'=\sqrt{4\pi/a}\,{\bf E}$, in self-explaining notation), the equation (\ref{2.140}) can take the two forms
\begin{equation} \int_{S^3} {\bf A}\cdot {\bf B}\,\dif^3 r = \gamma a,\qquad \int_{S^3} {\bf C}\cdot {\bf E}\,\dif^3 r = \gamma a,
\label{2.150b}\end{equation}
where $a=\hbar c \mu_0=3.97\times 10^{-32}\mbox{J}\cdot$ s/C.
The two quantities in (\ref{2.150b}) are called {\it magnetic helicity} $h_m$ and {\it electric helicity} $h_e$, respectively.

We define now the {\it electromagnetic helicity} $\cal H$, also called the {\it total helicity} or simply the {\it helicity}, of a field as the semisum ${\cal H}=(h_m+h_e)/2$
\begin{equation} {\cal H}=\frac{1}{2}\,\int_{S^3} (\A \cdot \B+{\bf C}\cdot \E)\dif ^3r=\gamma a.\label{2.170} \end{equation}
Note that the constant $a$, the helicity and the square of the Faraday 2-form have the same dimensions (action/charge)$^2$. Remind that, for simplicity, our choice of variables, defined in the Introduction, is $\lambda=1$ and $c=1$.

In order to understand better the idea of electromagnetic knots, we consider now an example with Hopf index $\gamma =1$,  linking number $\ell =1$ and helicity ${\cal H}=1$. To do that we need to know the values of the magnetic and electric fields at the initial time $t_0=0$. Let us define two maps $\phi, \theta: S^3\mapsto S^2$ so that we have the couple
\begin{equation} \phi (\R,0) =\phi_{\textrm H}(X,Y,Z)=\frac{(X+iY)}{(Z+i(R^2-1)/2},\,\quad \theta (\R,0)= \phi_{\textrm H}(Y,Z,X)=\frac{(Y+iZ)}{(X+i(R^2-1)/2}. \label{Q.10} \end{equation}
Because $ B_i=- \epsilon _{ijk}\Phi_{jk}(\phi)/2$ and $E_i=-\epsilon _{ijk}\Theta_{jk}(\theta)/2$ we can choose the following Cauchy data for the magnetic and electric fields
\beq \B(\r,0)=\frac{\sqrt{a}}{2\pi i}\,\frac{\nabla \phi \times \nabla \bar{\phi}}{(1+\bar{\phi}\phi)^2},\,\qquad \E(\r,0)=\frac{\sqrt{a}}{2\pi i}\,\frac{\nabla \theta \times \nabla \bar{\theta}}{(1+\bar{\theta}\theta)^2},\label{Q.20}\eeq
see (\ref{Tem.200}).
 From equations (\ref{Q.10}) and (\ref{Q.20}) it follows that
\begin{eqnarray} &&\B(\r, 0)= \frac{4\sqrt{a}}{\pi (1+R^2)^3}[2(Y-XZ),\,-2(X+YZ),\,-1-Z^2+X^2+Y^2]\label{P10},\\
&&\E(\r,0)=\frac{4\sqrt{a}}{\pi (1+R^2)^3}[1+X^2-Y^2-Z^2,\,-2(Z-XY),\,2(Y+XZ)]\label{P20}.\end{eqnarray}

It is easy to show that there are two vectors $\A$ and $\C$ that verify the  equations $\B=\nabla \times \A$ and $\E=\nabla \times \C$, so that
\begin{equation}\A(\r,0)=\frac{2\sqrt{a}}{\pi(1+R^2)^2}\,[Y,\,-X,\,-1],\quad
\C(\r,0)=\frac{2\sqrt{a}}{\pi(1+R^2)^2}\,[1,\,-Z,\,X].\label{P30}\end{equation}

 A question arises: could the vector $\A$ be understood as a vector potential of the theory?, or has it a different  meaning? To clarify this point is important, because if $\A$ is a genuine vector potential it could be applied to any formulation of the standard electromagnetism, while this would be impossible if this is not so.

We are interested here in the expressions  of $\B^2$, $\E^2$, $\A\cdot \B$ and $\C\cdot \E$. From equations (\ref{P10}), (\ref{P20}), (\ref{P30}), and after a bit of algebra, it easy to find that
\begin{equation}  \B^2=\E^2=\frac{16a}{\pi^2(1+R^2)^4},\qquad \A\cdot \B= \C\cdot\E= \frac{8a}{\pi^2(1+R^2)^4}\label{P100}\end{equation}
and \begin{equation} \A\cdot \B =\,\medio \B^2,\quad \mbox{ and }\quad \C\cdot \,\E =\frac{1}{2}\, \E^2  \label{P110}\end{equation}
so that \beq \A\cdot \B-\C\cdot \E=\frac{1}{2}(\B^2-\E^2)=0.\label{P120}\eeq
This indicates that this theory refers to the radiation fields.

It is important to understand properly which is the right meaning of the helicity 
\begin{equation} {\cal H}=\frac{1}{2}\,\int_{S^3} (\A \cdot \B+{\bf C}\cdot \E)\,\dif ^3r=\gamma a.\label{P130} \end{equation}
An equivalent form of (\ref{P130}) is the following equation for the electromagnetic energy
\begin{equation} {\cal E}=\frac{1}{4}\,\int_{S^3} (\B^2+\E^2)\,\dif ^3r=\gamma a.\label{P140} \end{equation}
These two equation are a consequence of Whitehead's theorem, see section 3.

\bs

\paragraph{Final comments.} We show here that there exist topological solutions of the electromagnetic field with basis in Hopf fibrations. Although they are classical, these solutions have a discrete nature. In the same way that General Relativity extended the dynamics to include the Geometry, changing thus the perspective of physics, the time is ripe to explore the resources of the topology.

\noindent {\bf Acknowledgements}:\\
We are grateful to profs. J. M. Montesinos and J. L. Trueba for discussions.\\

\noindent {\bf Competing financial interest}:\\
The two authors of this paper declare that they have no potential conflict of interest when submitting this article.

\end{document}